\documentclass[superscriptaddress,amsmath,amssymb,aip,titlepage,reprint]{revtex4-1}%
\usepackage[utf8]{inputenc}
\usepackage{amsfonts}
\usepackage{graphicx}
\usepackage{float}
\usepackage{subfigure}
\usepackage{natbib}
\usepackage{lipsum}

\usepackage{color,soul} %highlight

\begin{document}

\title{Time resolved imaging of the non-linear bullet mode within an injection-locked nano-contact spin Hall nano-oscillator}
\author{T. M. Spicer}
\author{P. S. Keatley}
\affiliation{Department of Physics and Astronomy, University of Exeter, Exeter, Devon, EX4 4QL, UK}

\author{M. Dvornik}
\affiliation{Department of Physics, University of Gothenburg, 412 96 Gothenburg, Sweden}

\author{T. H. J. Loughran}
%\author{V. V. Kruglyak}
%\author{R. J. Hicken}
\affiliation{Department of Physics and Astronomy, University of Exeter, Exeter, Devon, EX4 4QL, UK}

\author{A. A. Awad}
\author{P. D\"{u}rrenfeld}
\author{A. Houshang}
\author{M. Ranjbar}
\affiliation{Department of Physics, University of Gothenburg, 412 96 Gothenburg, Sweden}

\author{J. \r{A}kerman}
\affiliation{Department of Physics, University of Gothenburg, 412 96 Gothenburg, Sweden}
\affiliation{Materials Physics, School of ICT, KTH-Royal Institute of Technology, Electrum 229, 164 40 Kista, Sweden}

\author{V. V. Kruglyak}
\author{R. J. Hicken}
\affiliation{Department of Physics and Astronomy, University of Exeter, Exeter, Devon, EX4 4QL, UK}

%\date{Last updated \today}
%\maketitle

\begin{abstract}
Time-resolved scanning Kerr microscopy (TRSKM) has been used to image precessional magnetization dynamics excited by a DC current within a nano-contact (NC) spin Hall nano-oscillator (SHNO). 
%The SHNO was formed from a 4 micron diameter Py(5 nm)/Pt(6 nm) mesa defined upon a $Al_2O_3$ substrate, with triangular Au(150 nm) \hl{NCs} overlaid.
Injection of a radio frequency (RF) current was used to phase lock the SHNO to the TRSKM. The out of plane magnetization was detected by means of the polar magneto optical Kerr effect (MOKE).  However, longitudinal MOKE images were dominated by an artifact arising from the edges of the Au NCs.  Time resolved imaging revealed the simultaneous excitation of a non-linear `bullet' mode at the centre of the device, once the DC current exceeded a threshold value, and ferromagnetic resonance (FMR) induced by the RF current. However, the FMR response observed for sub-critical DC current values exhibits an amplitude minimum at the centre, which is attributed to spreading of the RF spin current due to the reactance of the device structure.  This FMR response  can be subtracted to yield images of the bullet mode.  As the DC current is increased above threshold, the bullet mode appears to increase in size, suggesting increased translational motion.  The reduced spatial overlap of the bullet and FMR modes, and this putative translational motion, may impede the injection locking and contribute to the reduced locking range observed within NC-SHNO devices.  This illustrates a more general need to control the geometry of an injection-locked oscillator so that the autonomous dynamics of the oscillator exhibit strong spatial overlap with those resulting from the injected signal.
%Time-resolved scanning Kerr microscopy has been used to image large amplitude precessional magnetization dynamics within a spin Hall nano-oscillator (SHNO).  The SHNO was formed from a 4 micron diameter Py(5 nm)/Pt(6 nm) mesa defined upon a $Al_2O_3$ substrate, with triangular Au(150 nm) contacts overlaid. Injection of an RF current was used to phase lock the SHNO to the repetition rate of the laser system.  Time resolved imaging revealed that injection of DC current leads to excitation of a non-linear `bullet' mode with a clear threshold behaviour, that can be separated from the small amplitude Ferromagnetic resonance (FMR) induced by the RF current.  The out of plane magnetization component is readily detected by means of the polar magneto optical Kerr effect (MOKE).  However images obtained by means of longitudinal MOKE measurements are dominated by an artifact arising from the edges of the Au contacts.  Micromagnetic simulations suggest that the diameter of the bullet mode is significantly smaller than that of the focused laser spot.  Nevertheless, as the DC current is increased above the threshold value, the image of the bullet mode is found to increase in size, suggesting that the bullet becomes increasingly mobile and exhibits significant translational motion relative to the centre of the device.  
\end{abstract}

\maketitle

%\section{Introduction}

Within a spin torque oscillator (STO), magnetic auto-oscillations, with MHz to GHz frequencies, are driven by the spin transfer torque (STT) associated with injection of DC spin current. Their frequency and amplitude can be tuned via either the DC electrical bias current or an applied magnetic field, while the magnetoresistance of the constituent materials leads to the generation of voltage oscillations. STOs have strong potential for magnetic sensing, signal processing, and neurmorphic computing applications \cite{stiles2006spin,Karenowska2015}.  The ability to lock the frequency and phase of the STO to an injected RF signal is an important property within applications, while arrays of STOs promise increased output power through mutual synchronization.  However it is first necessary to understand the character of the underlying magnetization dynamics. More specifically, the dynamics excited by both DC and RF currents must be determined if the conditions required for phase-locking are to be fully understood.

Within a spin Hall nano-oscillator (SHNO) the Spin Hall effect (SHE) \cite{Kato2004, Hoffmann2013} drives a pure spin current from a heavy metal with large spin-orbit interaction into a ferromagnet layer \cite{Demidov2012,Brataas2012,Dumas2014}. The de-coupling of charge and spin currents opens up new device geometries, for example enabling exploitation of magnetic insulators\cite{Hamadeh2014}, and in the present study, allows optical access to the active region of the device. %In SHNOs the local injection of spin current compensates non-linear damping and allows for formation of a localized non-linear mode dubbed the spin wave 'bullet'\cite{}.

The generation of  magnetic auto-oscillations requires a critical spin current density to be exceeded. Within the SHNO the injected charge current is concentrated  within a small region of the heavy metal, either by overlaying thick needle-shaped nano-contacts (NCs) on the heavy metal layer ,\cite{Demidov2012,Demidov2014,Liu2013,Nouriddine2011,Ulrichs2014,Giordano2014} or by forming a nanoconstriction within the heavy metal/ ferromagnet bilayer\cite{Demidov2014a,Awad2016,Kendziorczyk2016,Mazraati2016,Zahedinejad2017,Divinskiy2017}. SHNOs of both kinds have been studied by Brillouin Light Spectroscopy (BLS), microwave spectroscopy and micro-magnetic simulations. While the spectral characteristics of the dynamics have been explored, the time-dependent magnetization has not been measured directly. 

\begin{figure}
\centering
\includegraphics[width=8cm]{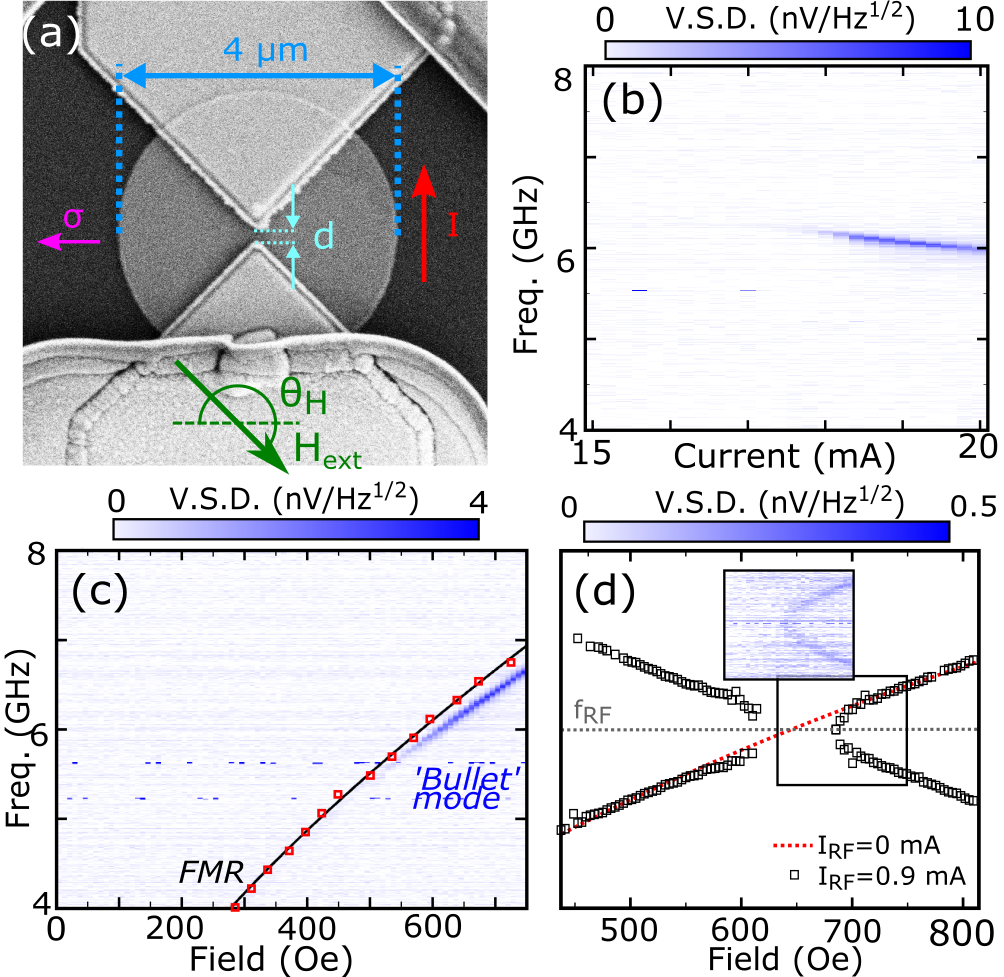}
\caption{(a) SEM image of a typical SHNO, where $I$ is the injected current, $\sigma$ the corresponding spin polarization, $d$ the NC separation, and $H$ the magnetic field applied at angle $\theta_H$. (b) Voltage Spectral Density (VSD) of microwave emission from a SHNO with $d$ = 240 nm at fixed magnetic field  $H$ = 650 Oe and $\theta_H = 210^\circ$ for different values of $I_{DC}$. (c) Microwave emission from a SHNO with $d =$ 180 nm for $I_{DC}$ = 18 mA, with magnetic field $H$ oriented at $\theta_H = 150^\circ$.  The red squares show resonance fields determined by STT-FMR measurements of the same device, while the solid black curve is a fit that is described within the main text. (d) Emission from a device with d = 200 nm when $I_{DC}$ = 16 mA and $I_{RF}$ has amplitude of 0.9 mA and frequency of 6 GHz. Squares identify the centre of peaks fitted to the emission spectra. The red dotted curve represents the line centre for free-running oscillations when $I_{RF}$ = 0 mA. The inset shows the emission spectrum at the edge of the locking region, from which a back-ground spectrum acquired with $I_{DC}$ = 0 has been subtracted.}%17dBm
\label{Fig:1}
\end{figure}

Here time resolved scanning Kerr microscopy (TRSKM)\cite{Gangmei2011,Valkass2016,Keatley2017,Keatley2016} is used to  study NC-SHNO devices. The response to both a radio frequency (RF) current $I_{RF}$, and a DC current $I_{DC}$  when phase-locked to an injected $I_{RF}$, are observed.  The source of the contrast observed in longitudinal and polar magneto optical Kerr effect (MOKE) measurements is first explained, before the spatial character of the magnetization dynamics induced by RF and DC currents is determined. Finally, the implications for effective injection locking and optimisation of the device geometry will be discussed.

%\section{Experimental method}

NC-SHNOs were fabricated by a combination of sputter deposition and electron-beam lithography. A 4$\mu m$ Py(5 nm)/Pt(6 nm) bi-layer disk was first defined upon a $Al_2O_3$ substrate, before two triangular Au(150 nm) NCs, with tip separation $d$, were overlaid as shown in Figure \ref{Fig:1}(a).  The NCs are intended to concentrate electrical current within the Pt at the NC tips. The charge current generates a spin current, via the SHE, that propagates from the Pt into the Py. Once the STT compensates the damping, a self-localized non-linear spin wave "bullet" mode is formed. While other modes can be supported within the disk, e.g. propagating waves when the Py is magnetized normal to the plane,\cite{Giordano2014} the bullet mode is of particular interest due to its narrow linewidth and tuneable frequency.

Initial microwave electrical measurements were performed by connecting a selected device to a bias-tee.  The inductive and capacitative arms were used to supply $I_{DC}$ and $I_{RF}$ respectively, while the RF signal from the device was directed into a spectrum analyzer via a circulator and +24 dB pre-amplifier. Stroboscopic TRSKM measurements were performed with a vector quadrant bridge detector that exploits different magneto optical Kerr effect (MOKE) geometries to simultaneously detect the three spatial components of the dynamic magnetization and the optical reflectance \cite{Valkass2016a}. The dynamics must be synchronized, via the injected $I_{RF}$, to an exact multiple of the 80 MHz repetition rate of the laser within the TRSKM. The phase of $I_{RF}$ is then adjusted relative to the laser pulses so that the time evolution of the magnetization dynamics can be observed. Measurements were performed with phase modulation of $I_{RF}$ to enhance the signal to noise ratio. The laser pulses had 800 nm wavelength, and were focused to a spot of $\sim$870 nm FWHM diameter by a microscope objective with 10.1 mm working distance and 0.55 numerical aperture \cite{Keatley2017}.

%\section{Electrical Measurements}
Electrical measurements were performed to identify the bullet mode and confirm locking to $I_{RF}$. Figure \ref{Fig:1}(b) shows emission from a NC-SHNO with $d$ =  240 nm when $I_{DC}$ exceeds $\sim$18 mA.  The frequency red-shifts with increasing $I_{DC}$, while no emission is observed if the sign of either $H$ or $I_{DC}$ is changed, in agreement with the expected symmetry of the SHE. Figure \ref{Fig:1}(c) shows the field and frequency dependence of both the Ferromagnetic Resonance (FMR) $f$, determined from separate STT-FMR measurements (consistent with previous work \cite{Liu2011,Liu2013}), and the microwave emission from the same device with $d$ = 180 nm. The FMR frequency has been fitted to the well known formula $f  = (\gamma/2\pi)\sqrt{[H(H+4{\pi}M)]}$ where $(\gamma/2\pi) = (g/2) \times$ 2.80 MHz/Oe with $g$ = 2, and $M$ = 604 $\pm$ 50 emu/cm$^3$. For a given frequency, microwave emission is observed at a field close to but greater than that of the FMR mode. The dependence of the frequency upon $H$ and $I_{DC}$ is consistent with previous observations of the bullet mode \cite{Ulrichs2014}.

Figure \ref{Fig:1}(d) shows locking of a bullet mode of 6 GHz frequency to an injected $I_{RF}$ of the same frequency for a device with $d$ = 200 nm.  As $H$ is decreased from its maximum value, the frequency of emission is ``pulled'' towards 6 GHz, where it remains within the locking range. Due to the smaller signal amplitude, the pulling is not seen at the lower end of the locking range.  Within the locking range, increased output power and reduced linewdith occurs, although here it is masked by the residual injected $I_{RF}$ that reaches the spectrum analyzer. An intermodulation mode can also be observed, decreasing in frequency with increasing field. Locking could also be achieved by injection of $I_{RF}$ with frequency of 12 GHz (see supplementary material) but $I_{RF}$ $\sim$ 3.5 mA was required to achieve even a narrow locking range, which has been attributed to thermal noise enhanced by the spin current \cite{Demidov2014}. 
%An optically-detected STT-FMR study of similar devices \cite{Spicer2018a} showed that the frequency-dependent reactance of the device geometry causes spatial spreading of $I_{RF}$ that drives FMR within the extended disk.  The spreading does not affect the DC current but becomes more pronounced as the frequency of the RF current is increased.
Therefore TRSKM measurements were performed with the frequency of $I_{RF}$ set to 6 GHz so as to minimise the amplitude of $I_{RF}$ required to achieve locking.

%\section{TRSKM OBSERVATIONS}

%\subsection{Full magnetic response}

%\subsection{Subtracted data}

\begin{figure}
\centering
\includegraphics[width=8cm]{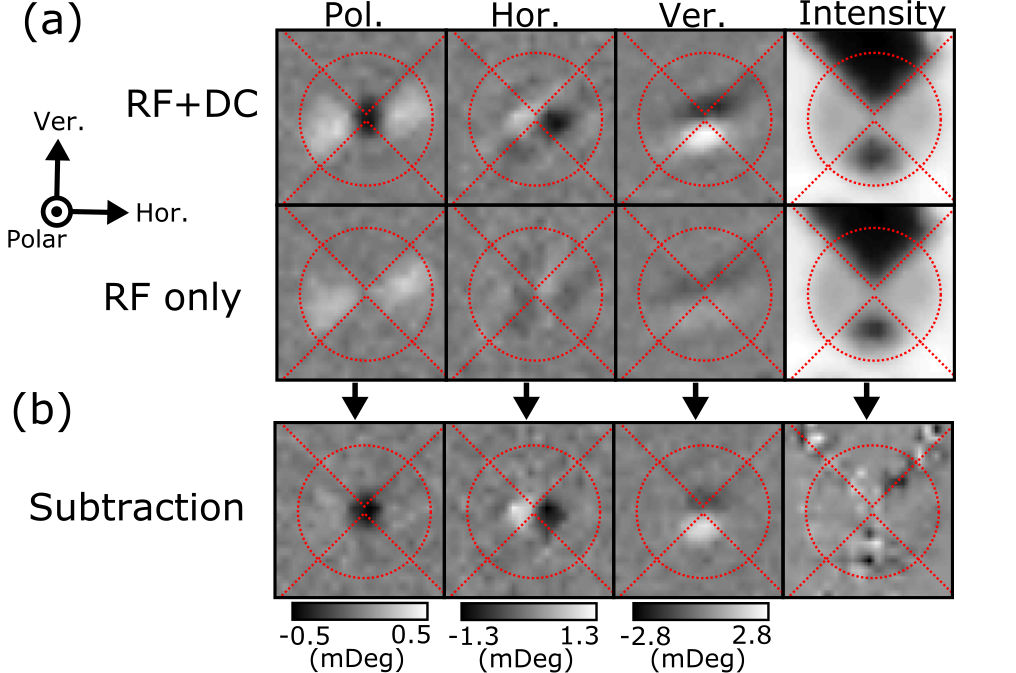}
\caption{(a) TRSKM images acquired from the polar (out of plane), horizontal, vertical and reflected intensity channels of the vector bridge detector for a NC-SHNO with $d$ = 200 nm,  $H$ = 650 Oe,  $\theta_H=210^\circ$, and $I_{RF}$ = 1.4 mA at 6 GHz frequency. $I_{DC}=$ 16 mA and 0 mA in the upper and lower panels respectively. (b) The difference of the upper and lower images from (a) is shown, revealing contrast due to the presence of $I_{DC}$ only. Each image shows a 5 $\mu m$ square region.}
\label{Fig:2} %9 dBm
\end{figure}

Figure \ref{Fig:2}(a) shows TRSKM images, of a device with $d$ = 200 nm, acquired with the bullet mode locked to $I_{RF}$, and when $I_{RF}$ is still present but $I_{DC}$ = 0 mA. In the latter case, $I_{RF}$ drives the FMR with $H$ detuned from the line centre. The addition of $I_{DC}$ leads to additional dynamics in all three magnetic channels. In the polar  channel, localized precession is observed between the NC tips, with a different phase to the dynamics in the extended disk (detailed within the supplementary material). By subtracting the images acquired with and without $I_{DC}$, the dynamic response due to $I_{DC}$ may be estimated, as shown in Figure \ref{Fig:2}(b). The subtracted images for the two in-plane (horizontal and vertical) channels each exhibit a spatially antisymmetric structure centred on the peak observed in the polar contrast, but occupying a somewhat larger area of $\sim 2\mu m$ diameter. The subtraction yields negligible residual contrast in the extended region of the disk, confirming that the bullet mode is tightly confined at the centre of the disk.

%The in-plane contrast does not conform to the expected bullet mode structure, making it difficult to identify it's origin. None contrast in figure \ref{Fig:2} was observed when the system was taken out of the injection-locking regime, ruling out mechanical motion.

Further measurements at different time delays confirmed that the contrast in the magnetic channels oscillates with $I_{RF}$, and with the same relative phase, which is unexpected if the magnetization undergoes a circular or elliptical precession. Furthermore, since $H$ is applied 30$^\circ$ from the horizontal axis, the amplitude of the dynamic magnetization is expected to be significantly greater in the vertical as compared to the horizontal direction, while in fact these two components were found to have comparable amplitude. To aid the interpretation, micromagnetic simulations were performed, using  MuMax 3 after the current distribution and associated Oersted field had been calculated in COMSOL \cite{COMSOL,Vansteenkiste2014}.  The magnetization of the permalloy disk was allowed to relax with $H =$ 1 kOe  and before the DC spin current and additional Oersted field were applied.

\begin{figure}
\centering
\includegraphics[width=8cm]{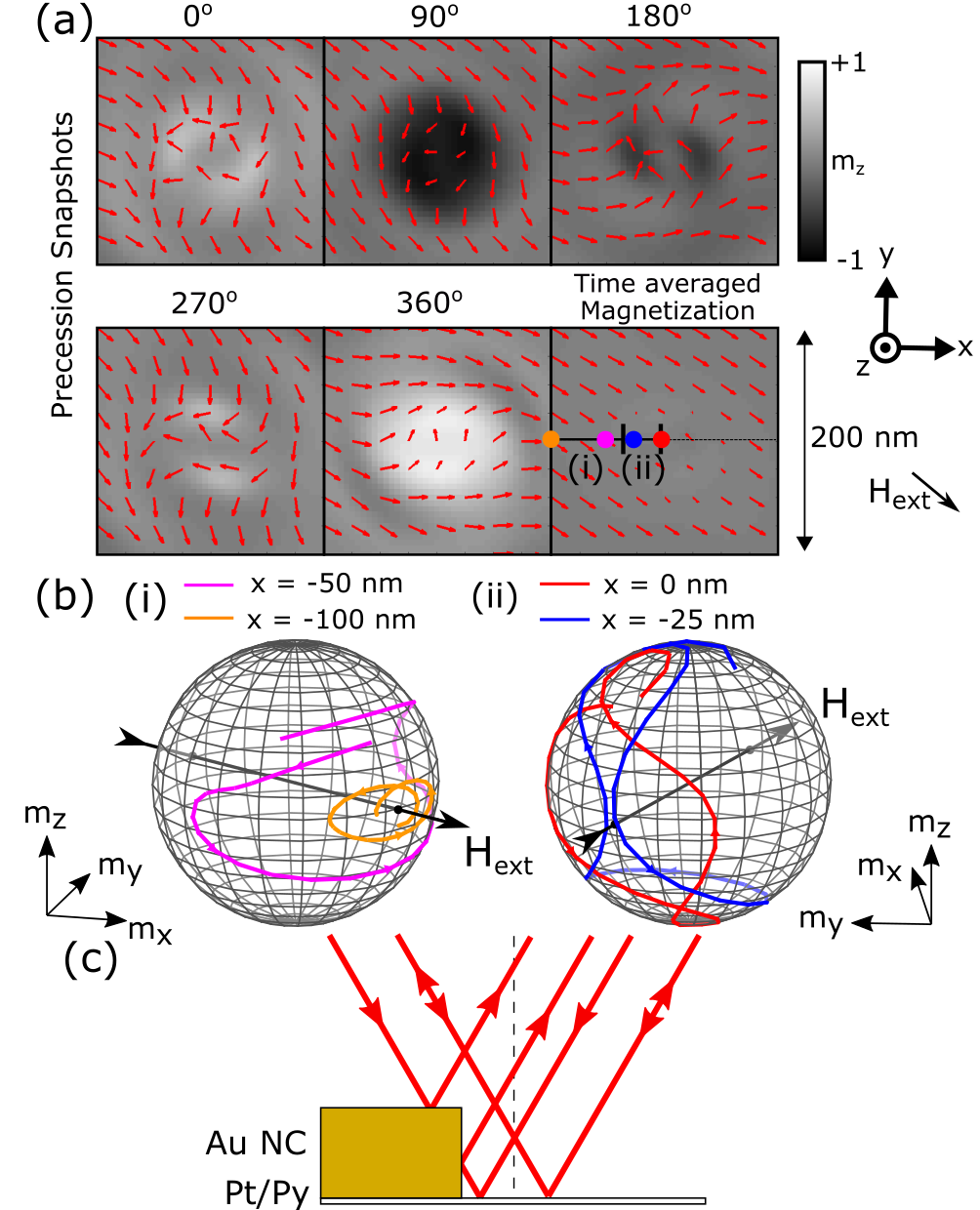}
\caption{(a) Simulated magnetization profile at different phase values within the cycle of auto-oscillation, with field $H$ = 1 kOe applied at $\theta_{H}$ = 210$^{\circ}$. The arrows indicate the projection of the magnetization within the plane, while the grayscale represents the out of plane component.  The images show the bullet mode with $\sim$ 70 nm diameter at the center of the device. The orientation of the coordinate axes is shown, while the origin lies at the center of the device. The lower right panel shows the time-averaged magnetization while the horizontal $x$ axis indicates two areas of interest, (i) the area immediately surrounding the bullet, and (ii) the core of the bullet mode. (b) Magnetization trajectories are shown for different points on the $x$ axis. Arrows show the direction of precession over 1 cycle of oscillation. (c) Schematic to illustrate the artefact that generates contrast in the horizontal and vertical channels. Arrows indicate the direction of propagation of rays.}
\label{Fig:3}
\end{figure}

Images of the simulated bullet mode are presented in Figure \ref{Fig:3}(a) for the configuration in Figure \ref{Fig:2}. Initial simulations showed that the bullet quickly escapes the active area and is damped within the extended disk. Therefore a pinning site was introduced in the form of either a single cell discontinuity in the magnetization, or a reduction in saturation magnetization with Gaussian spatial profile of 5\% peak value and $\sim 240$ nm FWHM as in Figure \ref{Fig:3}(a). This led to a bullet mode that was stable for a finite range of $I_{DC}$ values, as well as an additional mode that was localized in the non-uniform Oersted field associated with the injected charge current. The latter mode lies at higher frequency than the bullet mode, but was not observed in the room temperature electrical measurements of Figure \ref{Fig:1}, and so is not expected to appear in TRSKM measurements. Both the bullet and field-localized modes were found to have spatial and spectral character consistent with previous simulations \cite{Ulrichs2014,Giordano2014}. The need to pin the bullet may imply that the spin current produces a local reduction of the spontaneous magnetization that is not captured by the present simulations.

\begin{figure*}
\centering
\includegraphics[width=17cm]{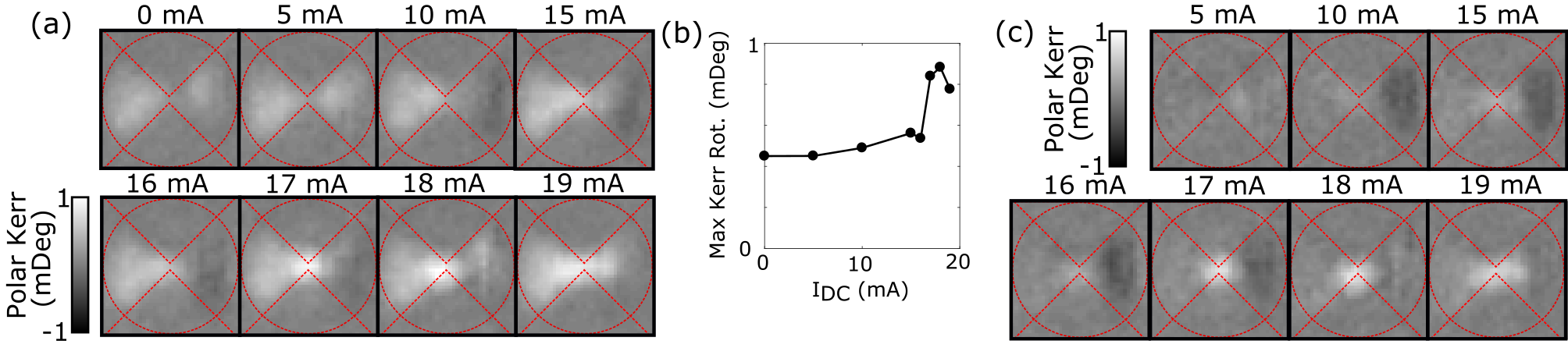}
\caption{(a) Polar TRSKM images acquired for different $I_{DC}$ values with the phase of $I_{RF}$ fixed. (b) Maximum absolute values of polar Kerr rotation extracted from the images in (a). (c) TRSKM images from (a) after subtraction of the $I_{DC}= 0$ mA image. All images were recorded from a NC-SHNO with $d$ = 240 nm, with $I_{RF}$ = 0.8 mA,  $H$ = 650 Oe  and $\theta_H = 210^\circ$.} %4dBm
\label{Fig:4}
\end{figure*}

While the bullet mode exhibits large angle precession, the images in Figure \ref{Fig:3}(a) do not reproduce the spatially antisymmetric character observed in the measured horizontal and vertical components. The core of the bullet mode undergoes the largest angle of precession. Figure \ref{Fig:3}(b) shows magnetization trajectories at different distances from the center of the disk along the $x$ axis, within two regions of interest, (i) outside and (ii) within the core of the bullet mode. The trajectories are plotted for one cycle of the bullet mode. They are not closed because the motion results from a superposition of the bullet mode with the field-localized mode that has a somewhat different frequency. Outside the core region the magnetization undergoes elliptical precession about an axis parallel to the applied field. At the edge of the core region, at $x=-35$ nm, the average magnetization is close to zero with an in-plane precession angle of $\sim270^\circ$. Within the core the precession amplitude increases further so that the  trajectory crosses over itself with the magnetization effectively precessing about a direction anti-parallel to the applied field. The magnetization precesses with the same phase at all positions within the disk. Simulations performed with an additional $I_{RF}$ demonstrated slightly improved stability of the bullet, but otherwise were of similar character.  

Difference images, calculated from simulated images separated by $180^\circ$ in phase, were convolved with a 870 nm FWHM Gaussian profile to more closely reproduce the experimental images.  Again they did not reproduce the spatially antisymmetric contrast observed in the vertical and horizontal channels. Further tests showed that the antisymmetric contrast was observed only when the bullet mode was present, ruling out mechanisms such as polarization of the Pt by $I_{RF}$ via the SHE.  Therefore the in-plane contrast must be an artefact associated with the optical probe overlapping the edge of the 150 nm thick NCs, while in proximity to the bullet mode. Figure \ref{Fig:3}(c) provides a schematic representation of the likely mechanism. As the probe passes over the NCs the beam returning to the detector is partially obstructed.  Crucially the symmetry between rays propagating in opposite directions within the cone is broken.  The resulting difference in intensity of the two halves of the back-reflected beam, combined with a finite polar Kerr rotation due to the bullet mode, manifests as a signal similar to that due to the longitudinal MOKE from an in-plane component of magnetization\cite{Keatley2006}. It follows from the NC geometry that a top-bottom antisymmetry is observed in the vertical channel and a left-right antisymmetry in the horizontal channel.

%\begin{figure}
%\includegraphics[width=8.5cm]{../images/fig4b_CurrVar_V3.eps}
%\caption{a) Polar TRSKM images acquired for different $I_{DC}$ values with the phase of $I_{RF}$ fixed. b) Maximum absolute values of polar Kerr rotation extracted from the images in (a) and also for the horizontal and vertical channels (not shown). (c) TRSKM images from (a) after subtraction of the $I_{DC}= 0$ mA image. All images were recorded from a SHNO with $d$ = 240 nm, with $I_{RF}$ = 0.8 mA,  $H$ = 650 Oe  and $\theta_H = 210^\circ$.} %4dBm
%\label{Fig:4}
%\end{figure}

The polar images are unaffected by the artefact. Figure \ref{Fig:4}(a) shows polar images, of a device with $d$ = 240 nm, acquired for different $I_{DC}$ values, with the phase and amplitude of $I_{RF}$ fixed. While the FMR mode is observed throughout the disk for all $I_{DC}$ values, the amplitude of the Kerr rotation at the NC tips is observed to increase markedly for current values in the range 17 to 19 mA. By extracting the maximum absolute Kerr rotation from Figure \ref{Fig:4}(a), a clear threshold behaviour, characteristic of the bullet mode, can be observed,  that is plotted in Figure \ref{Fig:4}(b). For small $I_{DC}$ values the amplitude of the FMR mode increases gradually with increasing $I_{DC}$ as the injection of DC spin current into the Py layer compensates the damping. In Figure \ref{Fig:4}(c) the images from Figure \ref{Fig:4}(a) have been replotted after subtracting the image for which $I_{DC}$ = 0. For $I_{DC}\geqslant 10$ mA a region of negative contrast appears to the right of the NCs. The asymmetry of the FMR response about the centre of the device reflects the mixed symmetry of the torques present.  The STT and the torque due to the in-plane Oersted field are symmetric about the centre while the torque due to the out of plane Oersted field is antisymmetric \cite{Spicer2018a}.

The electrical data of Figure \ref{Fig:1}(b) revealed the presence of a bullet mode for $I_{DC}$ values between 18 and 20 mA.  However Figure \ref{Fig:4} shows strong out of plane dynamics at the NC tips for $I_{DC}$ = 17 mA, that is still present when $I_{DC}$ = 19 mA. The reduced threshold value of $I_{DC}$ is due to the presence of the $I_{RF}$, as observed previously \cite{Demidov2014}.  Figures \ref{Fig:4}(a) and (c) also show that the apparent size of the bullet mode depends upon $I_{DC}$. Comparing the images for $I_{DC}$ values between 17 and 19 mA, the bullet mode appears to occupy a larger region as  $I_{DC}$ is increased, with some reduction in the maximum Kerr amplitude.  This might be explained by increased phase noise at the centre of the bullet suppressing its peak amplitude and causing its apparent width to increase.  However, since the diameter of this region is an order of magnitude greater than that of the simulated bullet mode in Figure {\ref{Fig:3}}(a), it seems more likely that the bullet instead develops significant translational motion when phase-locked to $I_{RF}$.

It is  clear from Figures \ref{Fig:2}(a) and \ref{Fig:4}(a) that the FMR induced by $I_{RF}$ exhibits a minimum at the centre of the device.  A more detailed optically-detected STT-FMR study has confirmed that the torques due to $I_{RF}$ also exhibit a minimum, which is attributed to lateral spreading of $I_{RF}$ due to the reactance of the device. \cite{Spicer2018a} The present study shows that the bullet and FMR modes exhibit weak spatial overlap.  It is reasonable to expect that the bullet may be pulled towards where the STT arising from $I_{RF}$ is larger, and may exhibit translational motion relative to the centre of the device.
The bullet could either establish a stable trajectory, or escape and be damped in the extended disk  (see supplementary material), allowing another bullet to form at the centre and repeat the process.  Increasing $I_{DC}$ is likely to increase the mobility of the bullet, allowing it to move further from the centre. 
Since the linewidth of the emission in Figure \ref{Fig:1}(b) is only weakly dependent on $I_{DC}$, formation of a stable trajectory seems more likely.  Both the lack of spatial overlap of dynamics induced by $I_{DC}$ and $I_{RF}$, and the translational motion of the bullet, may impede injection locking, and contribute to the NC-SHNO being difficult to lock \cite{Demidov2014} compared to other STO devices.

In summary, time resolved images of an injection-locked non-linear bullet mode within a NC-SHNO have been obtained.  The injected $I_{RF}$ excites a FMR mode that exhibits weak spatial overlap with the bullet mode. The apparent size of the bullet increases with DC current, which is suggested as being due to increased translational motion of the bullet when the RF current is present.  Further work is  required to determine the trajectory of the bullet. The translational motion and the lack of spatial overlap of the bullet and FMR modes may impeded injection-locking of the NC-SHNO.  This illustrates a more general need to control the geometry of injection-locked oscillators so that the autonomous dynamics of the oscillator exhibit strong spatial overlap with those resulting from the injected signal.
%\hl{These images confirm that the magnetization dynamics associated with the bullet are fully localized, so that mutual locking of multiple NC-SHNOs via a shared mesa seems infeasible}.  While the out of plane component of the dynamic magnetization was unambiguously observed, the in-plane components were obscured by an artefact due to the edges of the thick electrical contacts. 

\section*{Supplementary Material}
The supplementary material contains additional microwave spectroscopy and TRSKM datasets, and further details of micromagnetic simulations.

% use of bullets for data carriers: Karenowska2015}
\begin{acknowledgments}
We acknowledge financial support from the Engineering and Physical Sciences Research Council (EPSRC) of the United Kingdom, via the EPSRC Centre for Doctoral Training in Metamaterials (Grant No. EP/L015331/1) and EPSRC grants EP/I038470/1 and EP/P008550/1.

The research data supporting this publication are openly available from the University of Exeter's institutional repository at: https://doi.org/10.24378/exe.923
\end{acknowledgments}

%% Plasmon Notes:
%structure on scale of quarter wave is possible that a weak mode, not designed for plasmonics, at near tip not on scale smaller 1 um, a localized resonace would be on a smaller scale than 1 um, and a bulk mode would not likely be seen in both in-plane channels. Fabry-perot mode

%The effect of such interaction is unclear, if present it could enhance or interferer with the observed dynamics.

%un likely to couple into a strong plasonic mode, 

\bibliography{library_Adjust}{}  
\bibliographystyle{apsrev4-1}% prlrev4-1 unsrt apsrev4-1

\end{document}